# Distribution Models of Antennas in Radio Astronomy: Efficiency Comparison of the Golden Spiral Interferometry.

Author: Elio Quiroga Rodríguez
Universidad del Atlántico Medio, lecturer.
Las Palmas de Gran Canaria, Islas Canarias, España
elio.quiroga@pdi.atlanticomedio.es

**Abstract:**
This work compares the performance of different antenna configurations in radio astronomy interferometry, including the golden spiral, a grid, a random arrangement, and the "Y" configuration similar to the Very Large Array. One hundred antennas are simulated in each configuration, and the resulting UV coverage and image quality are analyzed. The results show that the golden spiral provides more uniform UV coverage without significant gaps, which improves image quality by reducing sidelobes and artifacts. In comparison, the grid exhibits periodic structures in the UV coverage that can degrade image quality due to gaps and artifacts. The random arrangement offers more natural coverage but is less efficient in terms of resolution and sidelobe control. The "Y" configuration proves effective in achieving high resolution along its arms but lacks complete coverage in certain directions, which can negatively affect image quality at those angles. The self-similar nature of the golden spiral allows for efficient capture of both large and small structures in observed sources, maximizing the spatial information obtained. We conclude that, for applications where resolution and sensitivity are critical, the golden spiral represents the optimal configuration, followed by the "Y" configuration, with the grid being the least suitable.

**Introduction and method:**
In a previous article[1], the author suggested using antenna configurations in interferometers based on the golden spiral, with promising results, indicating that a golden spiral distribution in a radio interferometer could significantly improve both resolution and sensitivity. This would be due to several factors, such as a more uniform distribution of radio telescopes and a longer maximum baseline. These improvements could lead to new discoveries in radio astronomy, such as the study of smaller or more distant objects. Additionally, a dynamic spiral interferometer would allow astronomers to optimize resolution, track motion, and study the structure of objects by adjusting antenna positions in real time.

However, comparing some of the most commonly used antenna distributions with the golden spiral proposal is still necessary to provide further support for the hypothesis. This article makes that comparison between four possible configurations, aiming to determine the pros and cons of each in order to draw conclusions.

**Discussion:**

A direct comparison between this interferometric array structure and commonly used alternatives appears to be necessary. Figure 1 shows the antenna distribution for four different configurations: the golden spiral, a square grid[2], a random arrangement to simulate a non-systematic distribution (a preferred model for authors such as Frédéric Boone[34]), and an Y shape[5], similar to the layout used in the Very Large Array[6]. These configurations will be used in a simulation to compare performance using 100 antennas. The Python code for this simulation is provided in the Annex of this text.

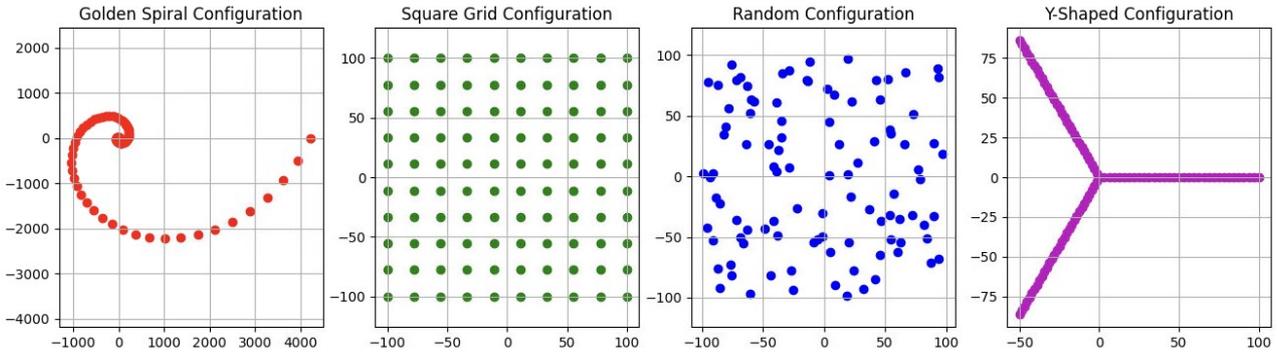

*Figure 1. Antennae distribution in a golden spiral, a square grid configuration, a random configuration and an "Y" shaped configuration (Author).*

In Figure 2, we can see a representation of UV-Coverage of the radio telescope resolution obtained with the three antennae distribution shown in Figure 1.

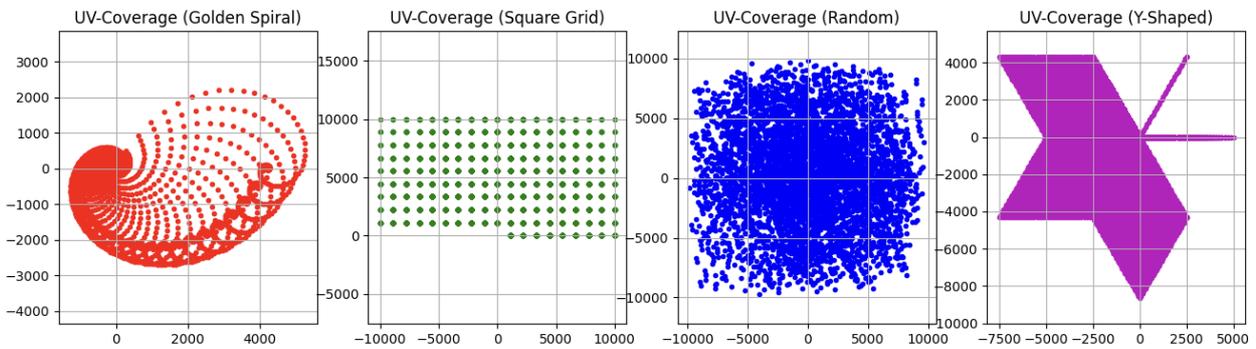

*Figure 2. UV-Coverage for radio telescopes with antennae distributions shown in Figure 1 (Author).*

Finally, Figure 3 shows a simulated image based on the UV-coverage from Figure 2 for each antenna distribution. As observed, the golden spiral distribution likely offers better UV-coverage with fewer gaps and smoother imaging properties due to its logarithmic nature. The square grid may display periodic structures in the UV-coverage, leading to sidelobes in the image, periodic gaps that may introduce significant artifacts and degrade overall imaging quality; the random configuration, while providing more natural coverage, may be less efficient in terms of resolution and sidelobe control. About the "Y" shaped configuration, it offers good resolution and sensitivity along the arms due to the long baselines. The configuration is well-suited for imaging elongated structures, particularly in the central regions, but tends to be sparse along certain directions perpendicular to the arms, which may lead to incomplete coverage and potential imaging artifacts in those directions.

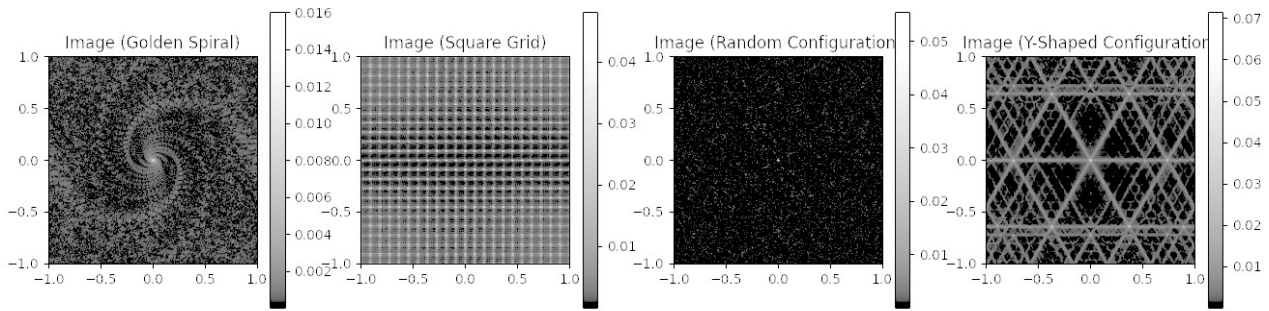

*Figure 3. Image simulation from UV-Coverage for radio telescope optinos shown in Figure 2; Golden Spiral, Square Grid and Random Configuration (Author).*

In terms of overall performance, including UV-coverage, resolution, and imaging quality, the configurations can be ranked roughly from best to worst as follows:

1. Golden Spiral: Offers a good balance between central and peripheral coverage, minimizing the risk of sidelobes and artifacts.

2. Y-Shaped configuration: Effective for achieving high resolution along the arms due to its long baselines, though its coverage can be sparse in certain directions.

3. Random Configuration: Due to its inherent randomness, its efficiency in terms of resolution and sidelobe control can vary, and the coverage may be uneven, potentially leading to lower image quality, contradicting Boone[7].

4. Square Grid: Although easy to implement and providing systematic coverage, it is prone to periodic structures in the UV-coverage, leading to sidelobes, gaps that can introduce significant artifacts, and degraded overall imaging quality, making it the least effective configuration of the four."

A more complex analysis falls outside the scope of this text. The author encourages the readers to perform their own.

The golden spiral's self-similar nature seems to maintain a good coverage at multiple scales, enabling the telescope to be sensitive to both large and small-scale structures in the observed objects. Additionally, this configuration minimizes redundancy in baseline lengths, maximizing the efficiency of the array by capturing a wider range of spatial information with a given number of antennas. The spiral pattern also provides a smooth transition between short and long baselines, which may be beneficial for capturing a continuous range of angular scales in the observed astronomical sources.

**Conclusions:**
In this work, we have explored and compared different antenna distribution configurations for use in interferometric radio astronomy arrays. The golden spiral has proven to be superior in several key aspects, including UV coverage, image quality, and efficiency in capturing spatial information. Its distribution minimizes redundancy in baseline lengths and provides a smooth transition between observed spatial scales, making it advantageous for both large and small structures. In comparison, the "Y" configuration stands out for its ability to offer high resolution along its arms, but its lack of coverage in directions perpendicular to them limits its application in certain cases. On the other hand, the random arrangement provides more uniform coverage than the grid but does not achieve the same level of sidelobe control and may be less efficient in terms of resolution. Lastly, the grid is the least favorable configuration due to the inherent issues in its periodic structure, which result in significant gaps and artifacts in the resulting images. Overall, we conclude that the golden spiral is the most balanced option for maximizing image quality and resolution in an interferometric array. This design is not only efficient in UV coverage but also facilitates better detail capture in astronomical observations, making it the preferred choice for future developments in radio astronomy.

**Data availability statement:**
All data is included in the article.

**Annex 1:**
The Python code used to obtain Figures 1 to 3 in this article.

```
import numpy as np
import matplotlib.pyplot as plt

# Define the number of antennas and the area
num_antennas = 100
array_radius = 100.0  # meters

# 1. Golden Spiral Configuration
golden_ratio = (1 + np.sqrt(5)) / 2
a = 10.0  # Initial radius
theta = np.linspace(0, 4 * np.pi, num_antennas)
r = a * golden_ratio ** theta
x_spiral = r * np.cos(theta)
y_spiral = r * np.sin(theta)

# 2. Square Grid Configuration
side = int(np.sqrt(num_antennas))
grid_x, grid_y = np.meshgrid(np.linspace(-array_radius, array_radius, side),
                np.linspace(-array_radius, array_radius, side))
x_grid = grid_x.flatten()[:num_antennas]
y_grid = grid_y.flatten()[:num_antennas]

# 3. Random Distribution Configuration
np.random.seed(42)
x_random = np.random.uniform(-array_radius, array_radius, num_antennas)
```

```python
y_random = np.random.uniform(-array_radius, array_radius, num_antennas)

# 4. Y-Shaped (VLA-like) Configuration
def y_shape_configuration(num_antennas, arm_length):
    # Distribute antennas equally along three arms
    antennas_per_arm = num_antennas // 3
    remaining_antennas = num_antennas - 3 * antennas_per_arm

    # Define angles for the Y shape (0°, 120°, 240°)
    angles = [0, 2 * np.pi / 3, 4 * np.pi / 3]
    x_y = []
    y_y = []

    for angle in angles:
        x_arm = np.linspace(0, arm_length, antennas_per_arm) * np.cos(angle)
        y_arm = np.linspace(0, arm_length, antennas_per_arm) * np.sin(angle)
        x_y.extend(x_arm)
        y_y.extend(y_arm)

    # If there are any remaining antennas, place them at the center
    x_y.extend([0] * remaining_antennas)
    y_y.extend([0] * remaining_antennas)

    return np.array(x_y), np.array(y_y)

# Use array_radius for the arm length
x_y, y_y = y_shape_configuration(num_antennas, array_radius)

# Function to calculate UV-coverage
def calculate_uv_coverage(x, y, wavelength=1.0):
    u = []
    v = []
    for i in range(len(x)):
        for j in range(i + 1, len(x)):
            u.append((x[j] - x[i]) / wavelength)
            v.append((y[j] - y[i]) / wavelength)
    return u, v

# Calculate UV-coverage for each configuration
u_spiral, v_spiral = calculate_uv_coverage(x_spiral, y_spiral)
u_grid, v_grid = calculate_uv_coverage(x_grid, y_grid)
u_random, v_random = calculate_uv_coverage(x_random, y_random)
u_y, v_y = calculate_uv_coverage(x_y, y_y)

# Plot Antenna Configurations
plt.figure(figsize=(16, 4))
plt.subplot(1, 4, 1)
plt.scatter(x_spiral, y_spiral, c='r', marker='o')
plt.title('Golden Spiral Configuration')
plt.axis('equal')
```

```python
plt.grid(True)

plt.subplot(1, 4, 2)
plt.scatter(x_grid, y_grid, c='g', marker='o')
plt.title('Square Grid Configuration')
plt.axis('equal')
plt.grid(True)

plt.subplot(1, 4, 3)
plt.scatter(x_random, y_random, c='b', marker='o')
plt.title('Random Configuration')
plt.axis('equal')
plt.grid(True)

plt.subplot(1, 4, 4)
plt.scatter(x_y, y_y, c='m', marker='o')
plt.title('Y-Shaped Configuration')
plt.axis('equal')
plt.grid(True)

plt.show()

# Plot UV-Coverage for each configuration
plt.figure(figsize=(16, 4))
plt.subplot(1, 4, 1)
plt.scatter(u_spiral, v_spiral, c='r', marker='.')
plt.title('UV-Coverage (Golden Spiral)')
plt.axis('equal')
plt.grid(True)

plt.subplot(1, 4, 2)
plt.scatter(u_grid, v_grid, c='g', marker='.')
plt.title('UV-Coverage (Square Grid)')
plt.axis('equal')
plt.grid(True)

plt.subplot(1, 4, 3)
plt.scatter(u_random, v_random, c='b', marker='.')
plt.title('UV-Coverage (Random)')
plt.axis('equal')
plt.grid(True)

plt.subplot(1, 4, 4)
plt.scatter(u_y, v_y, c='m', marker='.')
plt.title('UV-Coverage (Y-Shaped)')
plt.axis('equal')
plt.grid(True)

plt.show()
```

```python
# Create and compare images from UV-coverage
def create_image_from_uv(u, v, grid_size=256):
    uv_grid = np.zeros((grid_size, grid_size), dtype=complex)
    for ui, vi in zip(u, v):
        u_coord = int(ui + grid_size / 2)
        v_coord = int(vi + grid_size / 2)
        if 0 <= u_coord < grid_size and 0 <= v_coord < grid_size:
            uv_grid[u_coord, v_coord] += 1
    image = np.fft.ifftshift(np.fft.ifft2(np.fft.fftshift(uv_grid)))
    return np.abs(image)

image_spiral = create_image_from_uv(u_spiral, v_spiral)
image_grid = create_image_from_uv(u_grid, v_grid)
image_random = create_image_from_uv(u_random, v_random)
image_y = create_image_from_uv(u_y, v_y)

# Plot the resulting images
plt.figure(figsize=(16, 4))
plt.subplot(1, 4, 1)
plt.imshow(image_spiral, cmap='gray', extent=[-1, 1, -1, 1])
plt.title('Image (Golden Spiral)')
plt.colorbar()

plt.subplot(1, 4, 2)
plt.imshow(image_grid, cmap='gray', extent=[-1, 1, -1, 1])
plt.title('Image (Square Grid)')
plt.colorbar()

plt.subplot(1, 4, 3)
plt.imshow(image_random, cmap='gray', extent=[-1, 1, -1, 1])
plt.title('Image (Random Configuration)')
plt.colorbar()

plt.subplot(1, 4, 4)
plt.imshow(image_y, cmap='gray', extent=[-1, 1, -1, 1])
plt.title('Image (Y-Shaped Configuration)')
plt.colorbar()

plt.show()
```